\documentstyle[11pt,newpasp,twoside,epsf]{article}
\def\edcomment#1{\iffalse\marginpar{\raggedright\sl#1\/}\else\relax\fi}
\reversemarginpar

\begin{document}
\title{The double outburst of the unique object V838 Mon}
 \author{J.P. Osiwa{\l}a, M. Miko{\l}ajewski, T. Tomov, C. Ga{\l}an, J. Nirski}
\affil{Centre for Astronomy, Nicolaus Copernicus University, Pl 87100 Toru{\'n},
ul. Gagarina 11, Poland, sk@astri.uni.torun.pl,}
\author{D. Kolev, I. Iliev}     
\affil{NAO Rozhen, PBox 36, 4700 Smolyan, Bulgaria}
\begin{abstract}
We present spectroscopic and photometric observations of the recent 
peculiar outburst of V838 Mon, carried out at Rozhen and Toru\'n 
observatories. Our data cover a period of three months beginning 
just before the second eruption.
The evolution of the outburst is divided into four phases.
The changes of particular spectral features for each of these phases are
shortly discussed.
\end{abstract}
V838 Mon is one of the most enigmatic stellar phenomena observed during the last
several years. The outburst in the beginning of 2002 January (Brown 2002) with 
an amplitude in V of at least 5${^{\mathrm{m}}}$ pointed to a possible
classification of this object as a nova.
Its relatively cool K giant type spectrum almost unchanged during the
next three weeks (Zwitter \& Munari 2002) suggested that it could be an
extremely slow nova. Simultaneously, many metal lines showed P~Cyg profiles
with an outflow velocity of about 200--300 km s$^{-1}$, whereas the hydrogen 
Balmer lines presented
emission peaks without P~Cyg absorption components. 
Iijima (2002) suggested that we observed the flare-up phase of a post-AGB star. 
All these
speculations were broken about February~1, when an
unexpected second outburst (V amplitude $\sim$4${^{\mathrm{m}}}$) similar 
to the first one, occurred. Few days later, on about February 4--5, V838 Mon 
reached the maximum with almost 6\fm 5 in V~band. 


V838 Mon phenomenon has excited numerous observers, amateurs and
professional astronomers, which reflected in nearly 30
IAU Circular publications dedicated to this object. 
Additionally, almost 400 messages has been posted in the amateur VSNET network.
Recently, the first scientific
paper concerning V838 Mon has been published (Munari et al. 2002). 
However, there is no idea about the physical mechanism of V838 Mon
activity, yet.

We observed V838 Mon photometrically in $UBVR_cI_c$ system using the 60cm
Cassegrain telescope at Toru\'n Observatory with one-channel diaphragm
photometer equipped with a cooled C31034A photomultiplier. The high-resolution
(0.2~\AA/pix) spectra were collected with the coud\'e spectrograph 
fed by the 2m
telescope at Rozhen Observatory. Low-resolution spectroscopic observations
(2~\AA/pix) were carried out at Toru\'n Observatory using the Canadian
Copernicus Spectrograph attached at the Cassegrain focus
of the 90cm telescope. All our CCD spectra
were processed using the available tasks in the IRAF software package.
Part of the Toru\'n
spectra, obtained in the best weather conditions, were transformed to the
absolute 
\begin{figure}[t!]
\plotone{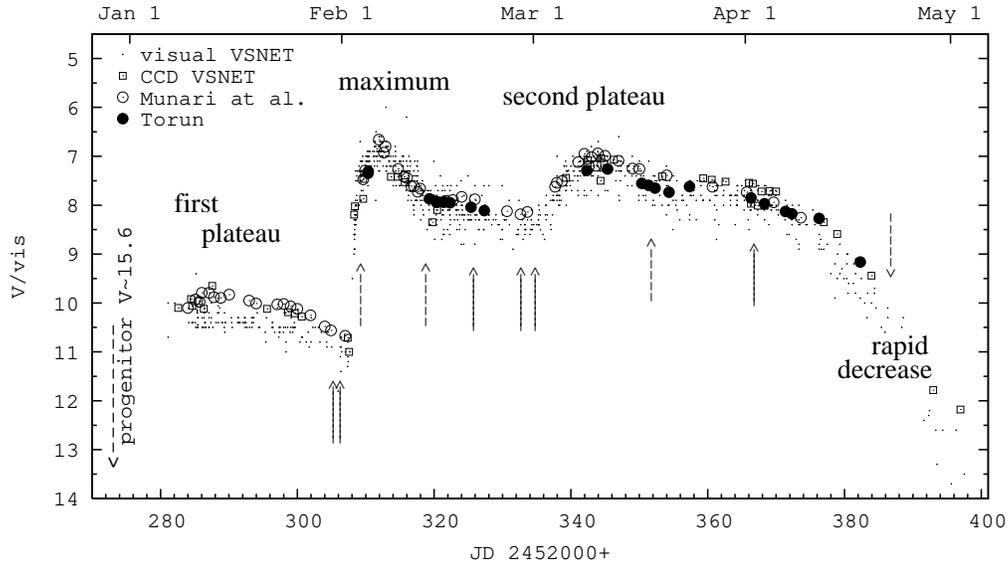}
\caption{Compiled V/vis light curve of V838 Mon during the 2002 outburst.
The dates when the spectra shown in Figs. 2 and 3 were obtained 
are denoted with solid (Rozhen) and dashed (Toru\'n) arrows.}
\end{figure}
\begin{figure}[t!]
\plottwo{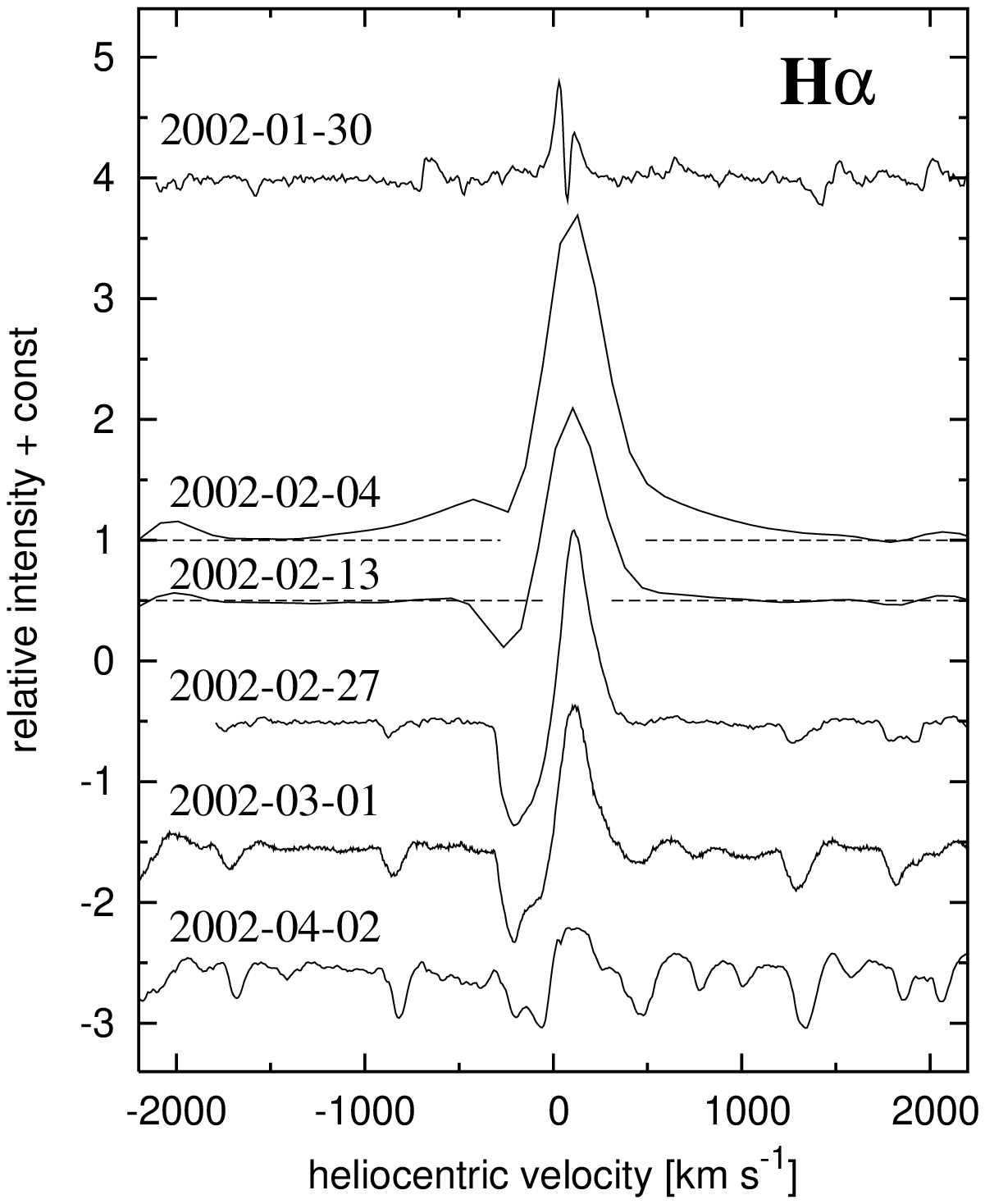}{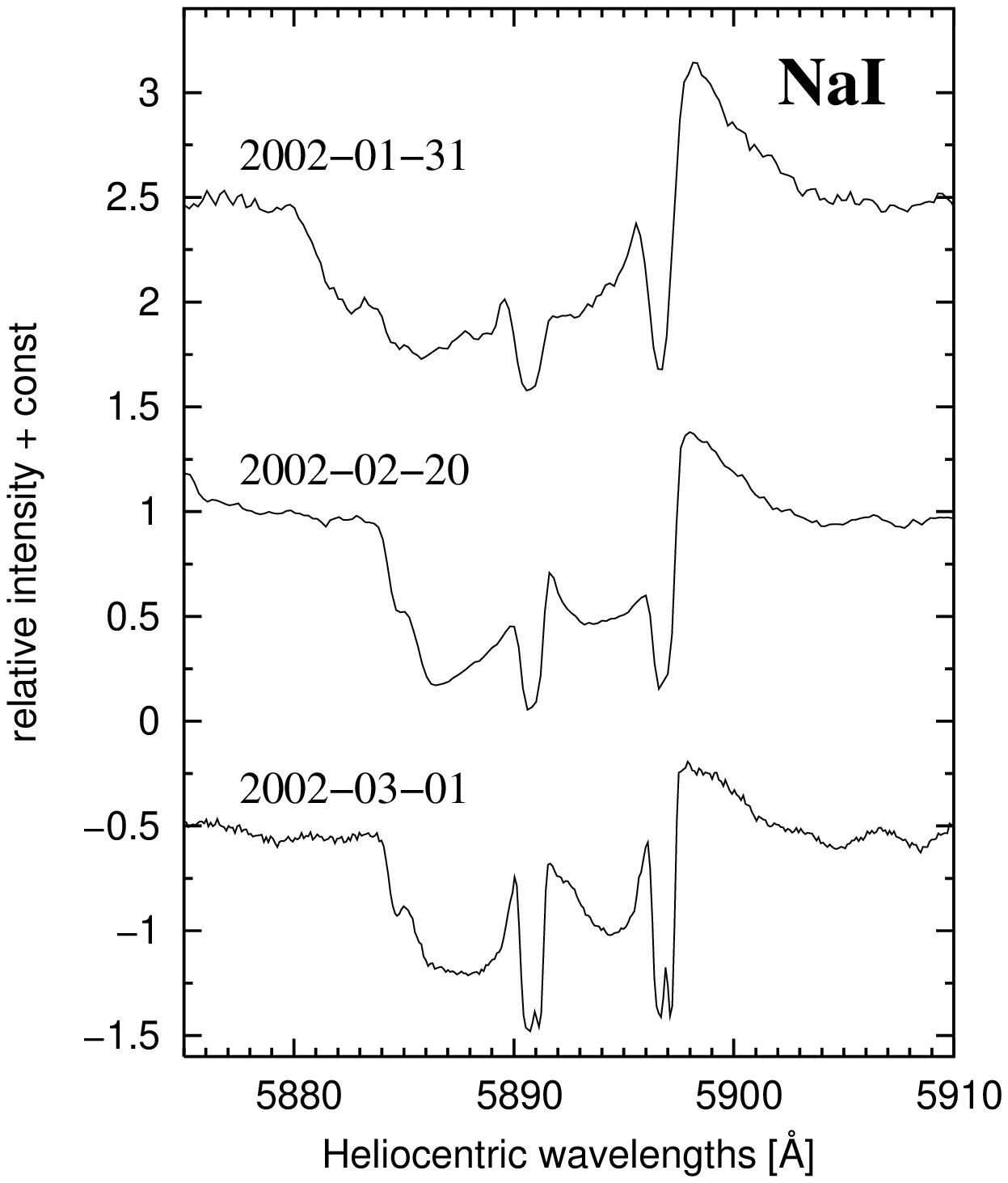}
\caption{Normalized to the local continuum profiles of H$\alpha$ (left) 
and NaI (right) lines. 
For two spectra in the left panel the local
continuum is shown with a dashed line to underline the remarkable variations
in the H$\alpha$ broad emission wings.}
\end{figure}
energetic scale (erg~cm$^{-2}$~s$^{-1}$~{\AA}$^{-1}$) by means of several
spectrophotometric standards observed during the same night.

The entire outburst of V838 Mon in 2002 lasted four months, 
from the beginning of January until the end of April. 
The brightness evolution of the object can be divided into four 
distinct periods (Fig. 1): 
\begin{itemize}
\item January -- first plateau between the first and second eruptions;
\item February 2--10 -- maximum of the brightness, most noticeable
      in U light;
\item February 11--April 15 -- second plateau;
\item after  April 15 -- rapid decrease in the optical range.   
\end{itemize}
The most interesting episodes during these periods are
illustrated by our spectra in Figs. 2 and 3.

During the period of the \emph{first plateau} several relatively strong P Cyg
profiles of low excitation and resonance lines of CaII, NaI, BaII, and others
were superimposed on a gK type (pseudo)photosphere spectrum.
As an example we show the profiles of Na I doublet 5890 {\AA}, 
5896 {\AA} (Fig.~2). 
The emission component of these lines indicates probably the stellar
velocity of about 50--70 km~s$^{-1}$.
The absorptions show outflow velocity about 200--300 km~s$^{-1}$ with a 
terminal one reaching $-$500 km~s$^{-1}$. 
At the same time H$\alpha$ shows relatively weak emission almost centrally
cut by a self-absorption component. The velocities measured for both of them
are about 50--70 km~s$^{-1}$, i.e. close to the stellar velocity as well.
A possible source of this
complex H$\alpha$ profiles could be a~stationary or rotating zone near the 
photosphere.

Our low resolution spectra obtained during the second outburst and its
\emph{maximum} suggested that the lines with low excitation potential practically
did not change, indicating that the region producing these features is
sufficiently far from the central object. At the same time 
 the flux in H$\alpha$ emission increased at least three times  
faster than the flux in continuum (Fig. 2).
Simultaneously,
we observed a rapid development of the P Cyg
absorption component. The most striking effect was the appearance of strong
and very broad H$\alpha$ emission wings extended to about $\pm$ 1600 km~s$^{-1}$. 
They have completely vanished very quickly after the second outburst. 
We did not
observe similar wings in any other lines in the spectrum of V838~Mon, and
what's more -- in other hydrogen lines. 
This argues for a non Doppler origin of the wings.
We suggest Raman scattering of Ly$\beta$ photons by neutral hydrogen as a
possible explanation. 
In the spectral region covered by our observations the best additional
indication for the presence of Ly$\beta$ photons (except H$\alpha$ wings!)
could be OI 8446 {\AA}
line excited by Bowen fluorescence (Bowen 1947). 
Unfortunately, we have obtained a spectrum in this region on February 21, 
when also the H$\alpha$ wings were at least one week
(i.e. since February 13, Fig. 2) completely invisible.
Nevertheless, a~very spectacular light-echo (e.g. Munari at al. 2002) 
observed around V838~Mon reveals
the high abundance of matter, dust and -- first of all -- neutral hydrogen
which must be a quantum trap for Ly$\beta$ photons. Moreover, the Doppler
structure of the Ly$\beta$ should be exactly the same as the main 
emission component of H$\alpha$, 
which shows the existence of an outflow
with escape velocity 200--300 km~s$^{-1}$. 
Raman scattering should multiply this
velocity by a factor $\sim$6.4 around H$\alpha$ wavelength, what corresponds
exactly to the width of the broad wings.

\begin{figure}[t!]
\plotone{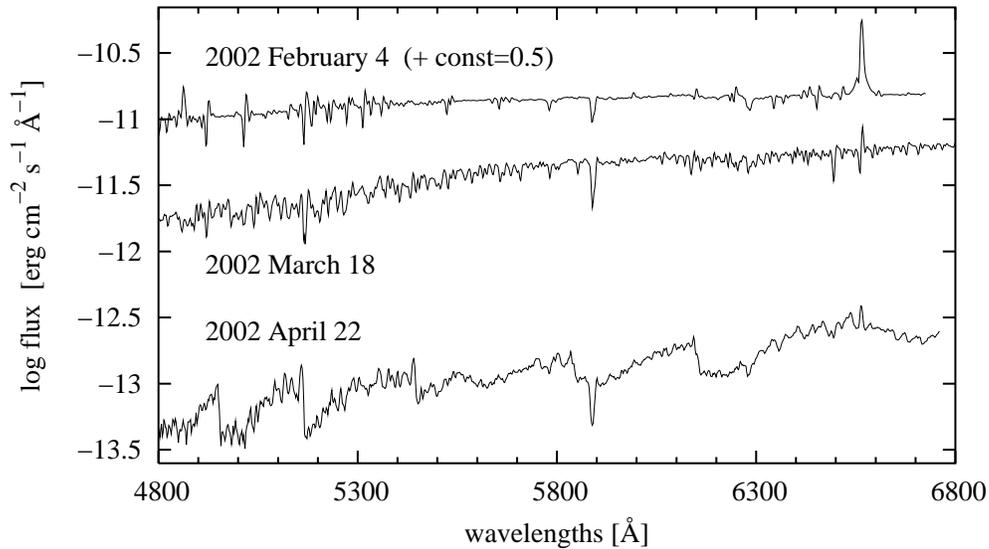}
\caption{The distribution of the energy in the low resolution spectra of
V838 Mon close to the 
\emph{maximum} (February 4), during the \emph{second plateau} (March 18) and
during the \emph{rapid decrease} (April 22).}
\end{figure}

During the \emph{second plateau} (pseudo)photosphere reaches again a gK
type, possibly luminosity class II (Fig.~3). H$\alpha$
showed pure P~Cyg profiles with emis-
\\
sion component decreasing in
intensity. The absorption component indicates a terminal velocity of nearly
300 km s$^{-1}$ (Fig.~2).  Also, the emission component of the low
excitation lines slowly weakened.  Remarkable decrease of the terminal
velocity is seen in the absorption component. The  P~Cyg absorption
component in NaI at the end of February was almost the same as in
H$\alpha$.

The rapid changes in the H$\alpha$ profile, observed during maximum
and the second plateau,
suggest the formation of an outflowing shell of matter, expanding with
velocity of about 250--300 km s$^{-1}$.

Finally, in the last \emph{rapid decrease} phase, surprisingly a pure M~giant
spectrum appeared.
In case the F0 subdwarf identified by Munari et al. (2002) is the true
progenitor and taking into account the lack of stellar wind the only way
V838 Mon to return to the progenitor state seems to be a contraction of the
present pseudophotosphere.

\acknowledgments
J. P. O. and T. T. are grateful to SOC and LOC for the support.
This study was supported by Polish KBN Grants No. 5 P03D~003~20 \& No.
5~P03D~013~21. We are grateful to the VSNET observers. 

\end{document}